\documentclass[doublecol]{epl2}

\usepackage{amsmath}

\graphicspath{{./figures/}}

\title{Hydrodynamics of Binary Fluid Mixtures---An Augmented Multiparticle Collison Dynamics Approach}
\shorttitle{Binary Fluid Mixtures}  

\author{Thomas Eisenstecken \and Raphael Hornung \and Roland G. Winkler \and Gerhard Gompper}

\institute{
Theoretical Soft Matter and Biophysics, Institute for Advanced Simulation and Institute of Complex Systems,
Forschungszentrum J\"ulich - 52425 J\"ulich, Germany
}

\abstract{
The Multiparticle Collision Dynamics technique (MPC) for hydrodynamics simulations is generalized to
binary fluid mixtures and multiphase flows, by coupling the particle-based fluid
dynamics to a  Ginzburg-Landau free-energy functional for phase-separating binary fluids.
To describe fluids with a non-ideal equation of state, an
additional density-dependent term is introduced. The new approach is verified by applying it
to thermodynamics  near the critical demixing point, and interface fluctuations
of droplets. The interfacial tension obtained from the
analysis of the capillary wave spectrum agrees well with the results based on the Laplace-Young
equation. Phase-separation dynamics follows the Lifshitz-Slyozov law.}

\begin{document}

\maketitle
\section{Introduction}

The description and modeling of multi-component and particle-laden flows is among the prime tasks
of fluid-dynamics approaches and simulations. As a consequence,  mesoscale
hydrodynamics simulation techniques have seen an enormous surge in interest,
development, and applications during the last decades. This was stimulated, on the one hand,
by the easy and straightforward coupling with nano- or micro-particles in the fluid, and
thus the possibility for the simulation of complex fluids, and,
on the other hand, the natural incorporation of thermal fluctuations, which are
essential, for example, for the conformations of flexible polymers under flow.

For the lattice Boltzmann method (LBM), an approach to simulate flows with multiple phases
and components has already been suggested in the early 1990s by Gunstensen et al. \cite{guns:91} and Shan and Chen
\cite{shan:93,shan:94}, and has been applied very successfully since then.
In this approach, the interaction between particles is described by a pseudopotential.
The method is only consistent with thermodynamic theories when the dependence of the
pseudopotential on the local density takes a special exponential form \cite{shan:94}.
This limits the accessible equations of state to be addressed.
An alternative approach has been suggested by Yeomans and collaborators a few
years later for non-ideal fluids \cite{swif:96} and for binary-fluid mixtures \cite{orla:95}.
This description is based on a free-energy functional, which enters the model via the
local pressure tensor. This approach has the advantage that any desired free-energy
functional can be employed.
For dissipative particle dynamics (DPD), the description of multi-component fluids
is rather straightforward, since there is an explicit conservative interaction between
particles, which is taken to be more repulsive for particles of unlike species \cite{cove:97}.

In contrast, the development of algorithms for multi-phase flows lags behind for a third
popular mesoscale simulation technique, multiparticle collision dynamics (MPC) \cite{kapr:08,gomp:09}.
This method has a priori no direct interactions between particles. Instead, it consists of alternating
streaming and collision steps. In the collision step, particles are sorted into the cells of
a simple cubic lattice. All particles in a given cell subsequently collide by exchanging momentum such
that the total momentum is conserved. This is achieved, for example, by a rotation of the particle
momenta in a comoving reference frame around a randomly oriented axis. This approach has been adapted to
multi-phase hydrodynamics by modifying the collision rule. This is achieved by a collision operator
that mimics an effective  repulsive interaction between different particles species
\cite{hash:00,inou:04,tuez:07,tuez:10,hill:16,eche:17}. For two-dimensional systems, the approach of
Ref.~\cite{tuez:07} is thermodynamically consistent and yields
the expected phase behavior, e.g., droplet formation \cite{tuez:10}. An effective repulsive interaction has also been employed recently
in a three-dimensional system to study
the flow of emulsion droplets in non–wetting microchannels, the apparent contact angle of sessile
droplets in mechanical equilibrium, and the dewetting of a liquid film from a wetting
stripe \cite{hill:16}, or the phase separation of binary fluids \cite{eche:17}. However, the latter
approach shares the difficulty of the pseudo-potential LBM that there is no inherent consistency
with thermodynamics.

Therefore, it would be highly desirable to have a free-energy based MPC method for non-ideal
fluids and fluid mixtures, in a similar spirit than the free-energy based LBM described above.
We develop such an approach in this paper, and verify its validity by applying it to phase
separation near a critical point, the Young-Laplace equation for fluid droplets, and the
interface profile between two coexisting phases. The advantage of the  method is its applicability
to a large variety of systems with different free-energy functionals.

\section{Model}

The MPC fluid is described by $N$ point particles of mass $m$, which move in continuous space with continuous velocities and interact with each other by a stochastic process \cite{male:99,kapr:08,gomp:09}.  The dynamics of the particles proceeds in alternating streaming and collision steps. For a bare MPC fluid, i.e., without an external force, the streaming motion is ballistic and the particle positions $\bm r_i$, $i=1,\ldots, N$, are updated according to
\begin{align}
\bm r_i (t+h) = \bm r_i(t) + h  \bm v_i(t) ,
\end{align}
where $h$ is the collision time and the $\bm v_i$ are the particle velocities.
Temperature is defined here by the average kinetic energy, i.e., $dNk_BT = \sum_{i=1}^N \langle mv_i^2 \rangle$ in
$d$ spatial dimensions; $k_B$ is the Boltzmann factor.
In the collision step, the system is partitioned into cubic cells of side length $a$, which define the local collision environment. Momentum exchange between the $N_c$  particles of a collision cell is achieved by a rotation of their relative velocities $\bm v_i (t) - \bm v_{cm}(t)$ around a randomly oriented axis by an angle $\alpha$, where $\bm v_{cm} = \sum_i \bm v_i /N_c$ is the center-of-mass velocity of the particles in the particular cell. The MPC-particle velocities are then given by
\begin{align}
\bm v_i(t+h) = \bm v_i (t) + (\mathbf{R}(\alpha)-\mathbf{E}) \left[\bm v_i (t) - \bm v_{cm}(t) \right] .
\end{align}
Here, $\mathbf{E}$ is the unit matrix and $\mathbf{R}$ denotes the rotation operator. The orientation of
the rotation axis is chosen randomly for every collision step and collision cell \cite{kapr:08,gomp:09}.
The algorithm conserves particle number, energy and local momentum, and thus gives rise to long-range
hydrodynamics \cite{huan:12}. The discretization implies a violation of Galilean invariance, which is
reestablished by the random shift of the collision-cell lattice \cite{ihle:01}.
It is important to emphasize that the MPC fluid has the equation of state of an ideal gas, i.e.,
$P_\text{mpc}= \langle N_c \rangle k_BT/ a^d$.

With a particle-based simulation approach, binary mixtures and their phase behavior can straightforwardly
be studied by introducing a suitable pair-wise interaction between the two types of particles. However,
such an approach would eliminate the main advantage of high computational efficiency of the MPC approach.
In order to preserve the simplicity of this algorithm, we couple the dynamics of the MPC particles to
the free-energy functional of a phase-separating system via the pressure tensor \cite{orla:95}.

We consider a binary-fluid MPC system and divide the $N$ particles into
$N_A$ particles of type $A$ and $N_B$ particles of type $B$.
Our starting point for the description of the interactions is the Ginzburg-Landau free-energy functional of a binary fluid,
\begin{equation}
\mathcal{F[\rho, \phi]}= \int	d^dr \left[ \frac{\tau}{2} \phi^2 + \frac{b}{4} \phi^4 + \frac{\kappa}{2}
                       \left(  \nabla \phi \right)^2 + g( \rho)  \right] \, ,
\label{FEnergy}
\end{equation}
where $\rho(\bm r)=\rho_A(\bm r) + \rho_B(\bm r)$ is the mass density and
$\phi (\bm r)= \rho_A (\bm r) - \rho_B (\bm r)$ the order parameter, with average
densities $\langle \rho_A \rangle=N_A/N$ and $\langle \rho_B \rangle=N_B/N$. Hence, our approach corresponds to Model H and belongs to the Ising universality class \cite{hohe:77, shim:15}. Phase separation is
determined by the parameter $\tau$, which (in the mean-field approximation) can be identified with  the
reduced temperature, i.e., $\tau=(T-T_c)/T_c$, where $T_c$ is the critical temperature.
The mass-density dependent term $g(\rho)$ is added to the free-energy functional to control the
fluid compressibility.
From Eq.~(\ref{FEnergy}), the pressure tensor $\mathcal{P}_{\alpha \beta}^{GL}$ can be derived
\cite{thei:99,orla:95,lamu:99},
\begin{equation}
\mathcal{P}_{\alpha \beta}^{GL} = \mathcal{P}_{\alpha \beta}^{b} + \mathcal{P}_{\alpha \beta}^{int}
                               + \mathcal{P}_{\alpha \beta}^{den}  , \label{differentP}
\end{equation}
where $\alpha, \beta \in \{x,y,z\}$ refer to the Cartesian components and
\begin{align}
\mathcal{P}_{\alpha \beta}^{b} &= \delta_{\alpha \beta} \left\{ \frac{\tau}{2}\phi^2
                                + \frac{3}{4} b \phi^4 \right\} , \label{firstP}\\
\mathcal{P}_{\alpha \beta}^{int} &= \delta_{\alpha \beta} \left\{ - \kappa \left[ \phi \nabla^2 \phi
             + \frac{1}{2} (\nabla \phi )^2 \right] \right\}
                          + \kappa (\nabla_{\alpha}\phi) (\nabla_{\beta} \phi), \label{secP}\\
\mathcal{P}_{\alpha \beta}^{den} &= \delta_{\alpha \beta}
                                 \left( \rho \frac{d g}{d \rho} - g \right) \label{thirdP}.
\end{align}
We denote the first term as ``bulk'' and the second as ``interfacial'' term. The third term is related to the
fluid equation of state and compressibility. In the following, we use $g(\rho)=\bar{\chi} \rho \ln (\rho)$,
which leads to an ideal-gas-like pressure tensor
$\mathcal{P}_{\alpha \beta}^{den}= \delta_{\alpha \beta} \bar{\chi} \rho$. The isothermal compressibility of the fluid is then given by $\kappa_T=1/(P_\text{mpc}+\bar{\chi} \rho)$. Hence, the parameter $\bar\chi$ can be varied to modify the fluid compressibility.
Alternatively, also a non-linear state function can be used, e.g., $g(\rho) = \chi \rho^2 /2 $, leading to the pressure tensor
\begin{equation} \label{eq:press_incomNON}
\mathcal{P}_{\alpha \beta}^{den}= \frac{1}{2} \delta_{\alpha \beta} \, \chi \rho^2 \, .
\end{equation}
Thereby, $\chi$ determines the strength of the fluid compressibility. We like to emphasize
that also other expressions for $g(\rho)$ can be employed for this purpose.

On the collision-cell level, the pressure gradients can  equally well be written as body-force density
$f_{\beta} = -\sum_\alpha \partial_{\alpha} \mathcal{P}_{\alpha \beta}$ acting on the fluid.
We discretise the field variables $\phi(\mathbf{r} )$ and $\rho(\mathbf{r} )$ on the collision-cell
level and in each of the MPC cells.
From Eqs.~(\ref{firstP})-(\ref{thirdP}), we can derive an expression for the force acting on the individual particles, which we split according to the three pressure contributions into
$F_{ \alpha} = F_{ \alpha}^{b} +  F_{ \alpha}^{int} + F_{ \alpha}^{den}$.
We discretize the bulk and compressibility
contributions $F_{ \alpha}^{b}$ and $F_{ \alpha}^{den}$ by averaging over two neighbouring cells.
As a consequence, while the scalar field variables $\phi$ and $\rho$ are defined in the collision cells,
the vector quantities $F_{\alpha}^{b}$ and $F_{\alpha}^{den}$ are defined on the edges of the collision
cells. As an example, the force in $x$-direction is given by
\begin{align} \label{discrete00}
F_{x+1/2, y}^{b}   = & -m \left[ \tau \left( \phi_{x+1, y} - \phi_{x, y} \right) \right. \\ \nonumber
& \left.   + \frac{3b}{2} \left( \phi_{x+1, y}^2 + \phi_{x, y}^2 \right) \left(  \phi_{x+1, y} - \phi_{x, y} \right) \right]  , \\ \label{discrete00_ideal}
F_{ x+1/2, y}^{den}   = &  -2 m \bar{\chi} (\rho_{x+1, y}-\rho_{x, y})/(\rho_{x+1, y} + \rho_{x, y} ).
\end{align}
Equation~(\ref{discrete00_ideal}) is the force due to an ideal-gas-like term. Similarly, the non-ideal-gas
expression can be used, with the discretization
$ F_{ x+1/2, y}^{den} = -m\,\chi \left( \rho_{x+1, y}-\rho_{x, y} \right)$.

For the interfacial contribution, $F_{ \alpha}^{int} $, we apply a simple central difference scheme, i.e.,  $\partial_{\alpha} \phi \approx \left( \phi_{\alpha+1} - \phi_{\alpha-1} \right) / 2$, which leads to
\begin{equation}
\begin{split}
F_{ x,y}^{int} = m \frac{\kappa}{2} [ \phi_{x+2, y} -\phi_{x-2, y} +\phi_{x+1, y-1} -\phi_{x-1, y-1} \\
 +\phi_{x+1, y+1}  - \phi_{x-1, y+1} -4\phi_{x+1, y} +4\phi_{x-1, y}   ] .
\end{split}
\end{equation}
The total forces in $x$-direction on particles of type $A$ and $B$ in cell ($x$,$y$) are then
\begin{align*}
F_{ x}^A & \!= F_{ x-1/2, y}^{b} + F_{ x+1/2, y}^{b}  +  F_{ x,y}^{int} + F_{ x+1/2, y}^{den} + F_{ x-1/2, y}^{den}, \\
F_{ x}^B & \! = -F_{ x-1/2, y}^{b} - F_{ x+1/2, y}^{b} -  F_{ x,y}^{int}+ F_{ x+1/2, y}^{den} + F_{ x-1/2, y}^{den}.
\end{align*}
The proposed discretization scheme guarantees that the sum of forces is globally equal to zero.
A more detailed derivation of the employed discretization scheme is presented in Appendix A. Finally, the modified streaming step is
\begin{align}
\bm v_{ i}(t+h) = & \ \bm v_{ i}(t) + \frac{\bm F_{ i}}{m}h , \\
\bm r_{ i}(t+h) = & \ \bm r_{ i}(t) + \bm v_{ i}(t)\, h + \frac{\bm F_{ i}}{2m}h^2,
\end{align}
where the  $\bm F_{ i}$ are given by the corresponding expressions for particles of type $A$ or $B$.

Temperature of the system is maintained by applying a cell-level canonical thermostat, where the relative velocities are scaled within a collision cell and the scaling factor is determined from the distribution of the kinetic energy in the cell~\cite{huan:10}.

\begin{figure}[t]
\begin{center}
\includegraphics[width=0.9\columnwidth]{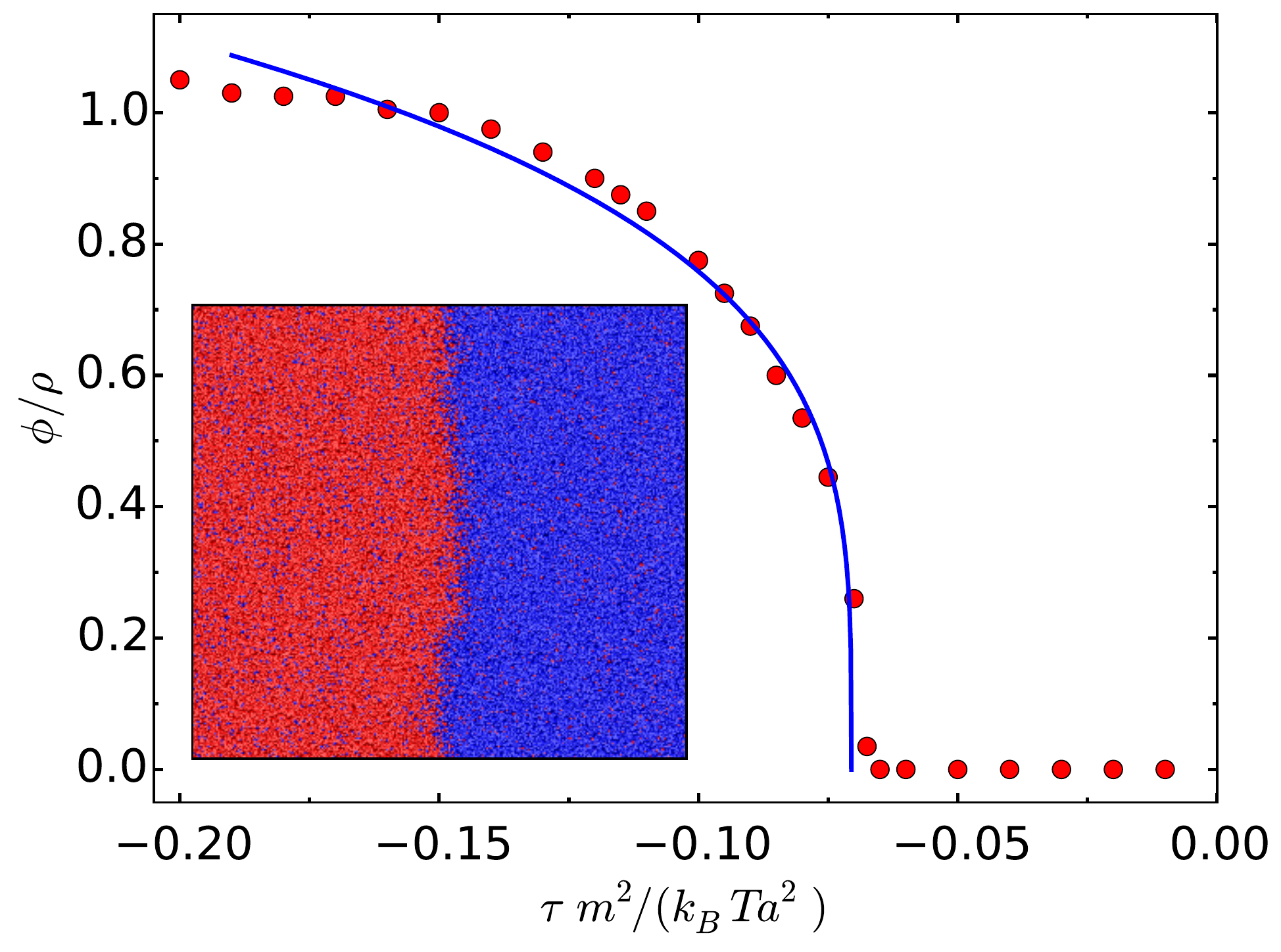}
  \caption{(Color online) Order parameter of a symmetric binary fluid as function of the reduced temperature $\tau$.
The symbols are simulation results and the line is a fit of Eq.~(\ref{eq:critical}) with
$\beta = 0.25 \pm 0.01$ and  $\tau_c \, m^{2}/( k_B T a^2) =-7\times 10^{-2}$.
Inset: Simulation snapshot for  $\tau \, m^{2} /(k_B T a^2) =-1.1\times 10^{-1}$.
The Ginzburg-Landau parameters are  $b  m^{4} /( k_B T a^6)=10^{-3}$, $\kappa m^{2}/(k_B T a^4)=0.3 $,
and $\bar{\chi} m/k_B T =5 $. }
  \label{phase}
\end{center}
\end{figure}

\begin{figure}[t]
\begin{center}
\includegraphics[width=0.9\columnwidth]{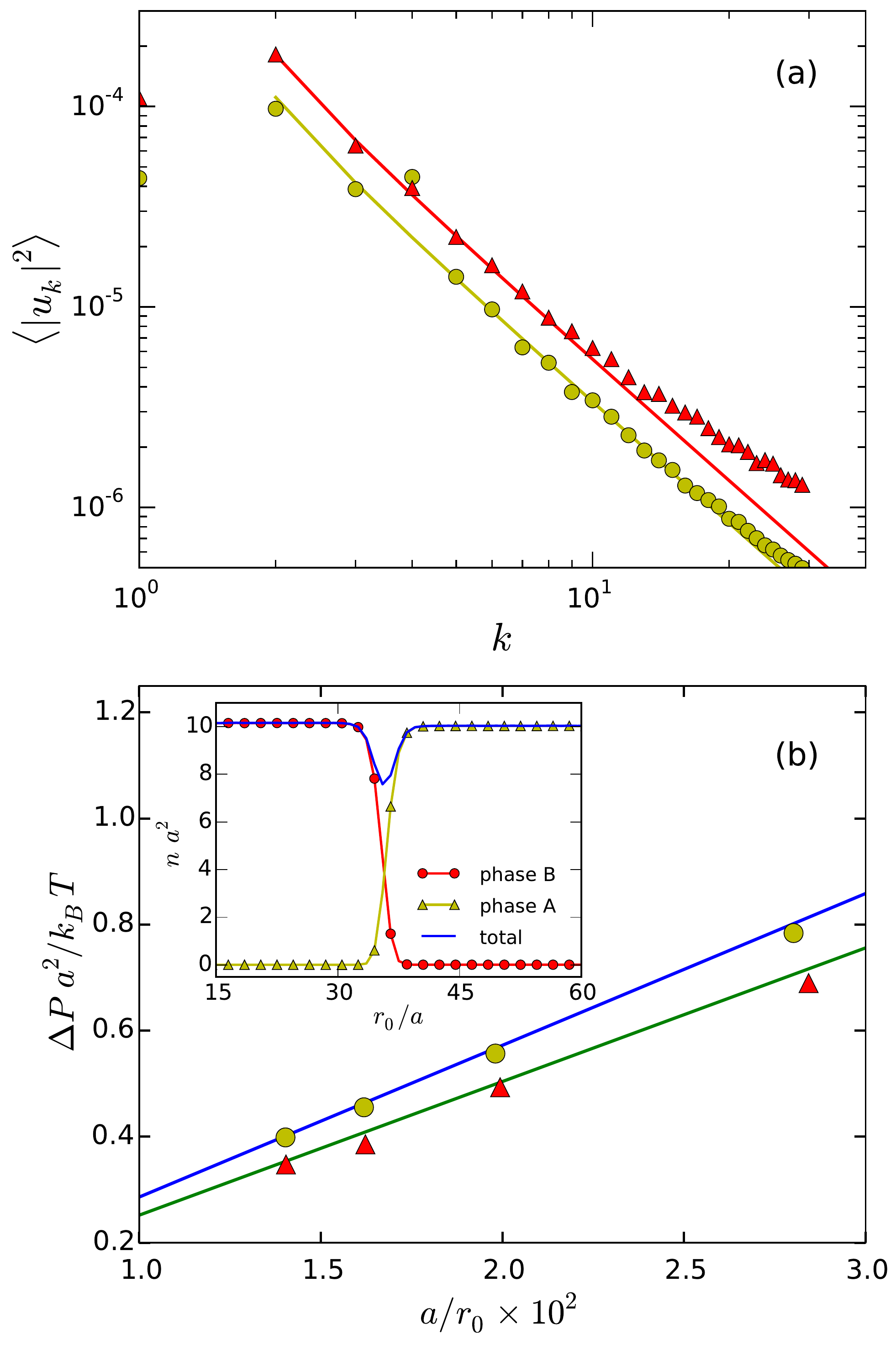}\label{capWave}
  \caption{(Color online) (a) Fourier amplitudes $\left<| u_k|^2 \right>$ of the radial thermal fluctuations of a
circular droplet (2D)  as a function of the mode number $k$ for $\tau m^{2}/(k_B T a^2) =-0.25 $
and $N_A/N_B = 3/2$ (yellow), as well as for $\tau m^{2}/(k_B T a^2) =-0.2$ and $N_A/N_B = 4$ (red).
The average droplet radii are $r_0/a=61.8 $ and $r_0/a=50.2 $. The solid lines are  fits to
Eq.~(\ref{fourAmp}).
(b) Pressure difference $\Delta P$ between the droplet and the surrounding solvent as a function of
the droplet radius $r_0$ for $\tau m^2/(k_B T a^2)=-0.25$ (yellow bullets) and
$\tau m^2/( k_B T a^2)=-0.2$ (red triangles). The lines are obtained with the line tensions following
from the corresponding capillary wave spectra (a). The inset shows the particle density as a function
of the radial distance from the droplet center of mass. In all plots,
$b\, m^4/(k_B T a^6)=10^{-3} $, $\kappa \, m^2/( k_B T a^4) = 0.3 $, and $\bar{\chi} \, m/k_B T =5$.}
\label{capWave_Pressure_plot}
\end{center}
\end{figure}

\section{Two-dimensional systems}

{\em Order parameter and critical exponent.}
Depending on the parameters of the bulk free-energy density in Eq.~(\ref{FEnergy}), the binary fluid
phase-separates into an $A$-rich and a $B$-rich phase (cf. Fig.~\ref{phase}).
As is well-known from the Landau (or mean-field) theory ($\phi=const.$), phase separation occurs for
$\tau<0$ (double-well potential) with the critical value at $\tau_c=0$. Since we are taking thermal
fluctuations and local field gradients into account, we expect a somewhat shifted value, with $\tau_c<0$.
In the vicinity of continuous phase transitions, the order parameter exhibits a power-law dependence
of the form \cite{land:69,reic:16}
\begin{equation} \label{eq:critical}
\phi \sim  \left(\tau-\tau_c \right)^\beta ,
\end{equation}
with critical exponent $\beta$ and critical-temperature shift  $\tau_c$.
Landau theory predicts the critical exponent $\beta = 1/2$, while the exact solution of the 2D Ising-model
yields $\beta = 1/8$ \cite{land:69,wu:82}.

To study the critical behavior of our A-B mixture, we employ a two-dimensional system with a box length
$L_x=L_y=100a$, and an average particle number per collision cell of $\langle N_c \rangle =10$. The measured temperature is increased by approximately $4\%$ as a consequence of the additional forces by the coupling of the MPC fluid to the order parameter field. Further, we set the
particle fractions to $N_A/N_B=1$. The collision step, as for all the following 2D simulations, is
performed with a rotation angle of $\alpha=90 ^{\circ}$ and the collision time
$h/\sqrt{ma^2/k_BT}=5\times 10^{-2} $.
For the phase diagram, the reduced-temperature parameter $\tau$ in the Ginzburg-Landau expression is
varied within the interval $\tau  m^2 / (k_B T a^2) \in [-0.2,\, 0]$. The order parameter is displayed
in Fig.~\ref{phase}. A fit of Eq.~(\ref{eq:critical}) yields the critical exponent
$\beta= 0.25 \pm 0.01$, in-between the Landau-theory and Ising-model values. We see two explanations
for this deviation. First, our model contains thermal fluctuations and correlations beyond mean-field
theory, so that Landau theory should not apply. Second, the true critical behavior occurs in a very
small temperature interval near the critical temperature, which we have not attempted to resolve,
since the purpose of our study is not a precise determination of critical exponents. Outside this
interval, a crossover to Landau behavior is expected \cite{anis:95}, which is consistent with our results.

\subsection{Interfacial tension of 2D droplets}
\label{ch:MPC_binMixInterfacial}

For the two Ginzburg-Landau parameters $\tau m^{2}/(k_B T a^2)=-0.25$, $-0.2$ and the particle fractions
$N_A/N_B = 3/2$  and $4$, respectively, the fluid phase separates and a circular droplet appears
(the simulation box size is $L_x=L_y=200 a$). To measure the interfacial tension of the  droplet of
radius $r_0$, we follow the approach of Refs.~\cite{tuez:07,tuez:10,stot:07} and determine the Fourier
amplitudes $u_k$ of the thermal fluctuations. According to the equipartition theorem, the amplitudes
are given by  \cite{tuez:10}
\begin{equation}
\langle | u_k|^2 \rangle = \frac{2k_B T}{\pi r_0 \sigma}\left( \frac{1}{k^2 - 1} \right)\, ,
\label{fourAmp}
\end{equation}
with line tension $\sigma$ and mode number $k$. Figure~\ref{capWave_Pressure_plot}(a) shows simulation
results for two reduced temperatures. For small mode numbers, the Fourier amplitudes  decay according
to Eq.~(\ref{fourAmp}). Thereby, the quadratic decay of $\langle |u_k|^2 \rangle$ for high $k$ is
more pronounced for the particle fraction $N_A/N_B = 3/2$ compared to $N_A/N_B = 4$. This can be
explained simply by the larger droplet radius for $N_A/N_B = 3/2$, which leads to a
better resolution of the interfacial undulations. However, a large droplet radius leads to an
interference with periodic images (due to the finite system size) and, hence, to a pronounced $k=4$
mode. This mode is less pronounced or even absent for the smaller droplet.
The fit of Eq.~(\ref{fourAmp}) yields the line tensions  $\sigma \approx 26.7 \, k_B T / a$
(for $\tau \, m^{2}/(k_B T a^2)=-0.25$) and $\sigma \approx 23.6 \, k_B T / a$
(for $\tau \, m^{2}/(k_B T a^2)=-0.2$). Thereby, for simulations with a pronounced $k=4$ mode, the
fit was taken in the interval $k\in [5, \, 15]$, and for simulations with a small droplet radius
for $k\in [2, \, 10]$.

Alternatively, the line tension can be calculated from the Young-Laplace equation \cite{ihle:06,tuez:07,tuez:10}
\begin{equation}
\Delta P = \sigma / r_0 \, . \label{eq:Young-lapl}
\end{equation}
The pressure difference $\Delta P = P_B -P_A $ for the fluid inside and outside of the droplet is measured via Eq.~(\ref{differentP}), where we
neglect the interfacial terms. This can be justified by measuring the pressure far away from
the interface. Additionally, the ideal-gas contribution arising from the particle motion of the MPC
algorithm, $P_\text{mpc}= \langle N_c \rangle k_BT/ a^2$, has to be taken into account. Simulations
with different droplet sizes are performed and the time-averaged values of $\rho$ and $\phi$ are
used to calculate the pressures $P_A$ and $P_B$, and $P_\text{mpc}$.
The pressure difference, as a function of the curvature $r_0^{-1}$ for the two values of the
reduced temperature $\tau$ is displayed in Fig.~\ref{capWave_Pressure_plot}(b). Evidently, the
Laplace-Young relation is well satisfied with the interfacial tension obtained from the
capillary-wave spectrum.  The inset of Fig.~\ref{capWave_Pressure_plot}(b) shows time-averaged radial
particle-density profiles for $N_A/N_B = 9$. Due to the repulsive interaction between the different
particles types, a minimum of the total density $\rho$ appears at the interface; the
equilibrium radius ($r_0$) of the droplet is taken to be at that point.
Figure~\ref{capWave_Pressure_plot}(b) illustrates a slight density difference between the droplet
and the surrounding fluid, arising from the self-compression of the droplet due to its surface (line)
tension.

\begin{figure}[t]
\begin{center}
\includegraphics[width=0.9\columnwidth]{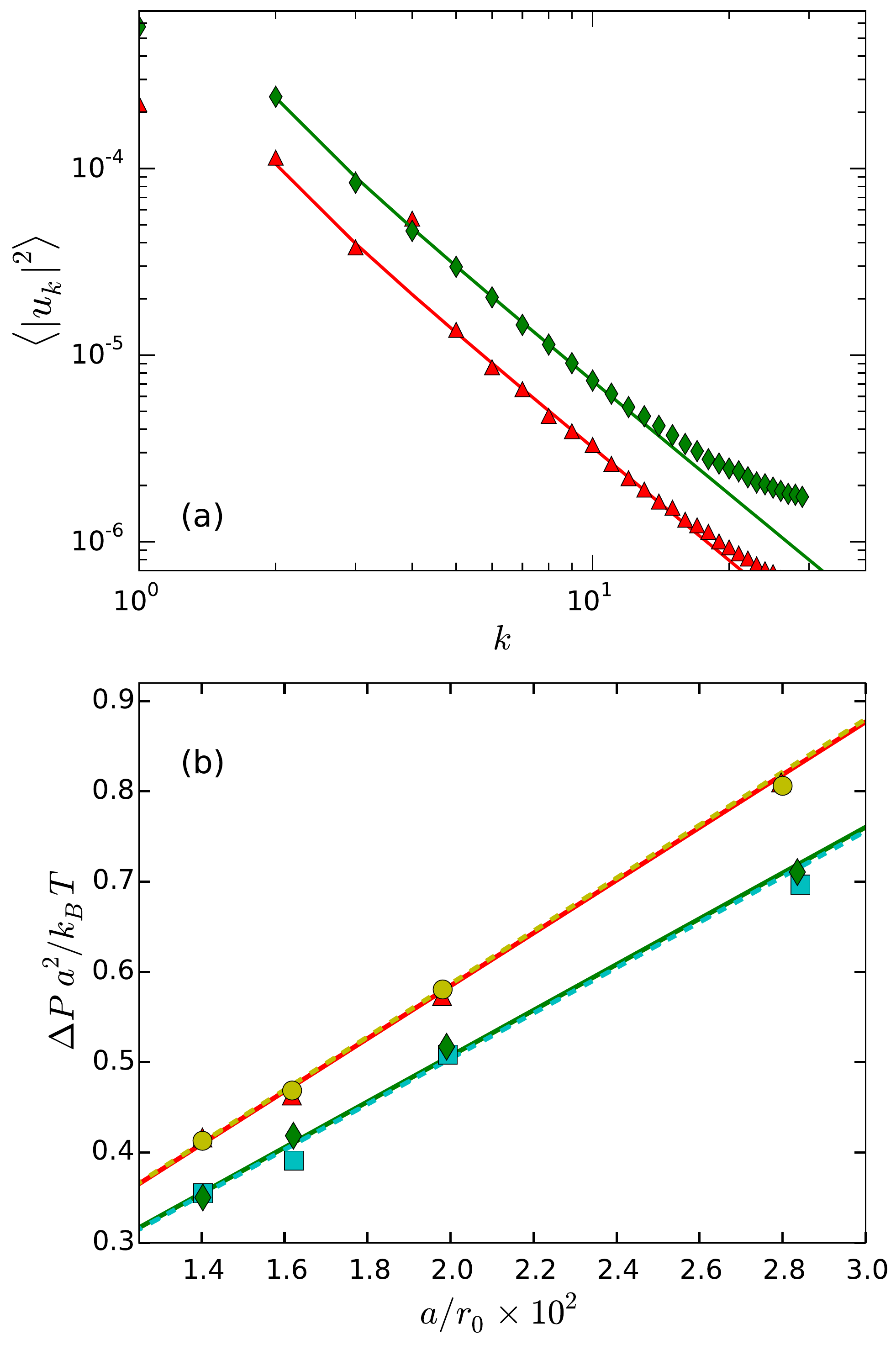}
  \caption{(Color online) (a) Fourier amplitudes  $\left<| u_k|^2 \right>$ of the radial thermal fluctuations of a
circular droplet (2D)  as a function of the mode number $k$ for $\tau m^{2}/(k_B T a^2) =-0.25 $ and
$N_A/N_B = 3/2$ (red), as well as for $\tau m^{2}/(k_B T a^2) =-0.2$ and $N_A/N_B = 9$ (green).
For both parameter sets,  $\chi \, m^2 /(k_B T a^2) =0.5$. The solid lines are fits to Eq.~(\ref{fourAmp}).
(b) Pressure difference $\Delta P$ between the droplet and the surrounding fluid as a function of
the droplet radius $r_0$ for the ideal-gas-like equation of state (dashed lines) and Eq.~(\ref{eq:press_incomNON}) (solid lines),
respectively, and the two values of $\tau$. The red triangles and yellow bullets correspond to
$\tau m^2/(k_B T a^2) =-0.25$, $\chi m^{2}/(k_BT a^2) =0.5$ (red), and $\bar{\chi} \, m/k_BT = 5$ (yellow).
The green diamonds and cyan squares correspond to $\tau\, m^2/(k_B T a^2) =-0.2$,
$\chi \, m^{2}/(k_BT a^2) =0.5$ (green), and $\bar{\chi} \, m/k_BT = 5$ (cyan). The lines illustrate the line tensions obtained from the corresponding capillary wave spectra.
For all plots, the other parameters are set to $b\, m^4/(k_B T a^6)=10^{-3} $ and
$\kappa \, m^2/( k_B T a^4) = 0.3 $.}
\label{Pressure_plot_nonId}
\end{center}
\end{figure}

\subsection{Nonlinear equation of state}
\label{ch:swarm_app_incompress}

As pointed out above, a non-ideal equation of state can be used to reduce compressibility effects,
for example as determined by the pressure tensor in Eq.~(\ref{eq:press_incomNON}).
In order to obtain comparable simulation results for systems with the different equations of state, we
choose the corresponding parameters $\chi$ and $\bar{\chi}$ such that the non-linear pressure contributions for both approaches are
approximately equal, i.e., we set $\chi=0.5$ and $\bar{\chi}=5$.

Figure \ref{Pressure_plot_nonId} displays the fluctuation spectra for two different reduced temperatures
$\tau$ and particle fractions $N_A/N_B$, as well as the pressure difference as function of the droplet
size. We obain again very good agreement with the theoretically expected power-law dependency of the
capillary wave spectrum (Eq.~(\ref{fourAmp})). Similarly, the Laplace pressure is well reproduced with
the line tension extracted from the fluctuation spectrum (Fig.~\ref{Pressure_plot_nonId}(b)).
This illustrates that our approach works well with different equations of state.

\section{Three-dimensional systems}

\begin{figure*}[t]
\begin{center}
  \includegraphics[width=0.275\textwidth]{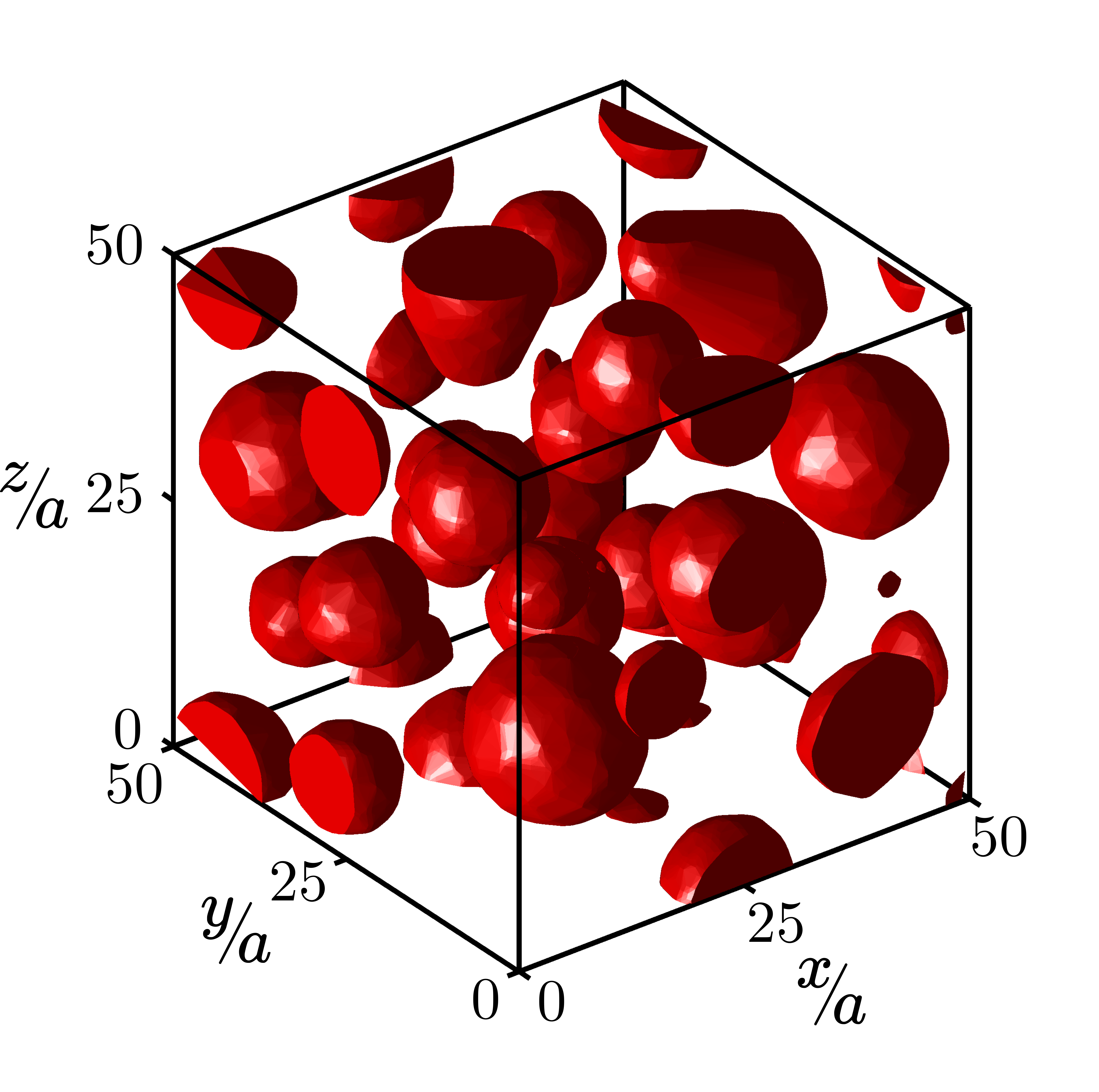}
  \includegraphics[width=0.275\textwidth]{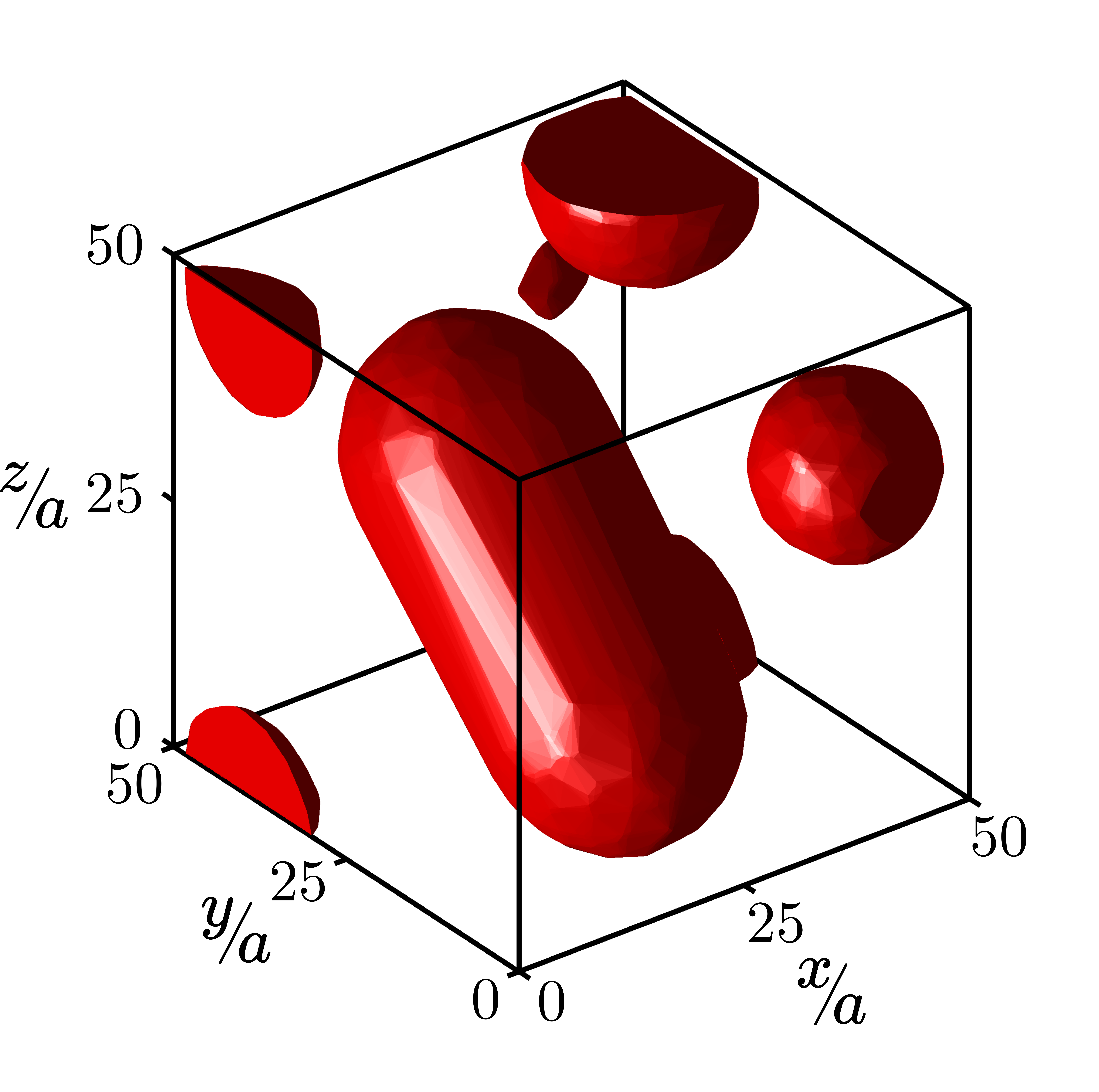}
  \includegraphics[width=0.275\textwidth]{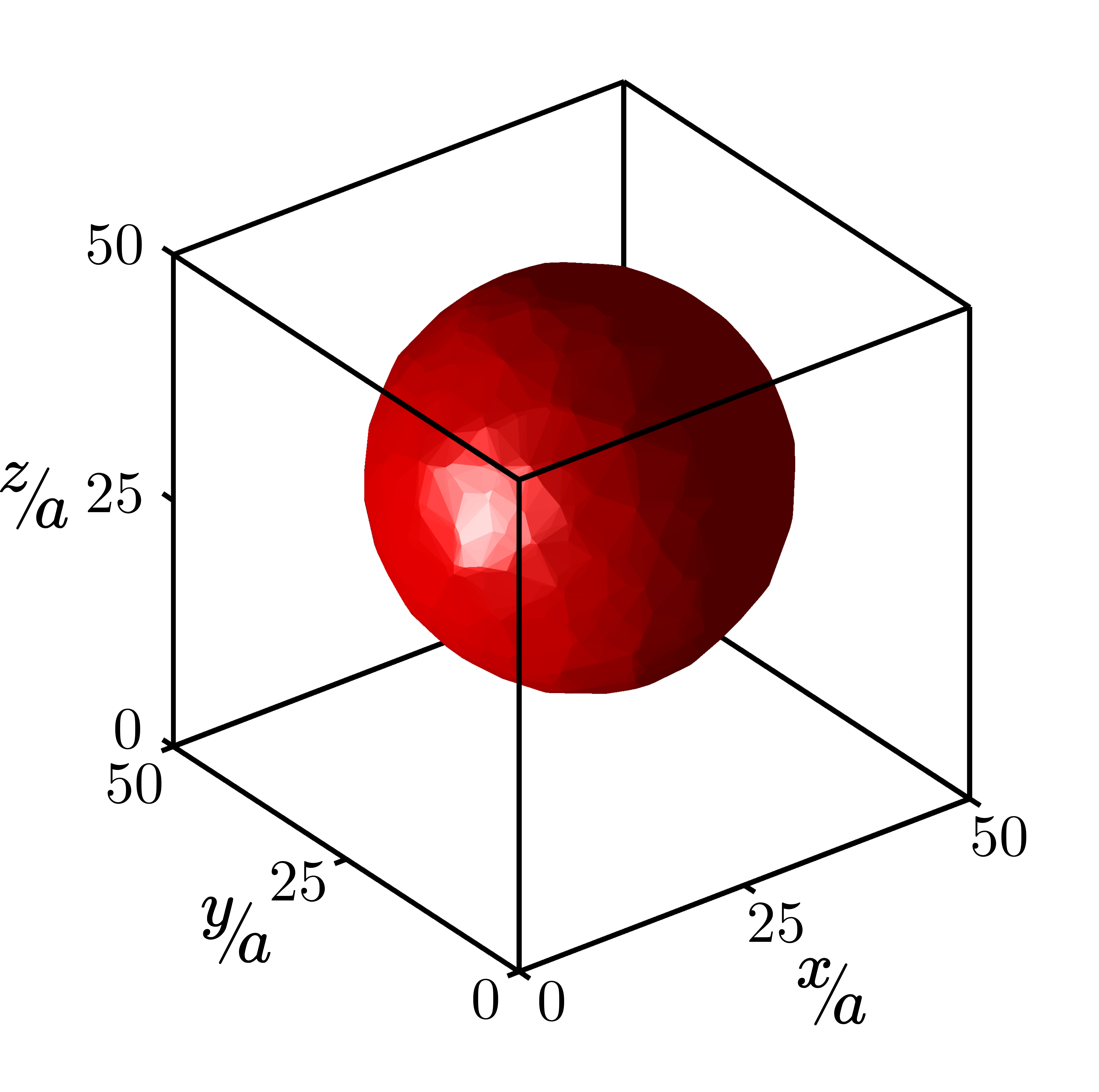}
\caption{(Color online) Snapshots of clusters of particles of type $B$ after $5 \times 10^3$, $6\times 10^4$, and $11\times 10^4$ simulation steps (left to right) for a cubic simulation box of size  $L_x=L_y=L_z=50a$.  The initially totally mixed fluid phase separates and converges to a minimal free-energy state. See also Supplemental Material {\em suppl.mov} for an animation of the droplet growth.}\label{3D_1}
\end{center}
\end{figure*}

\begin{figure}[t]
\begin{center}
\includegraphics[width=0.9\columnwidth]{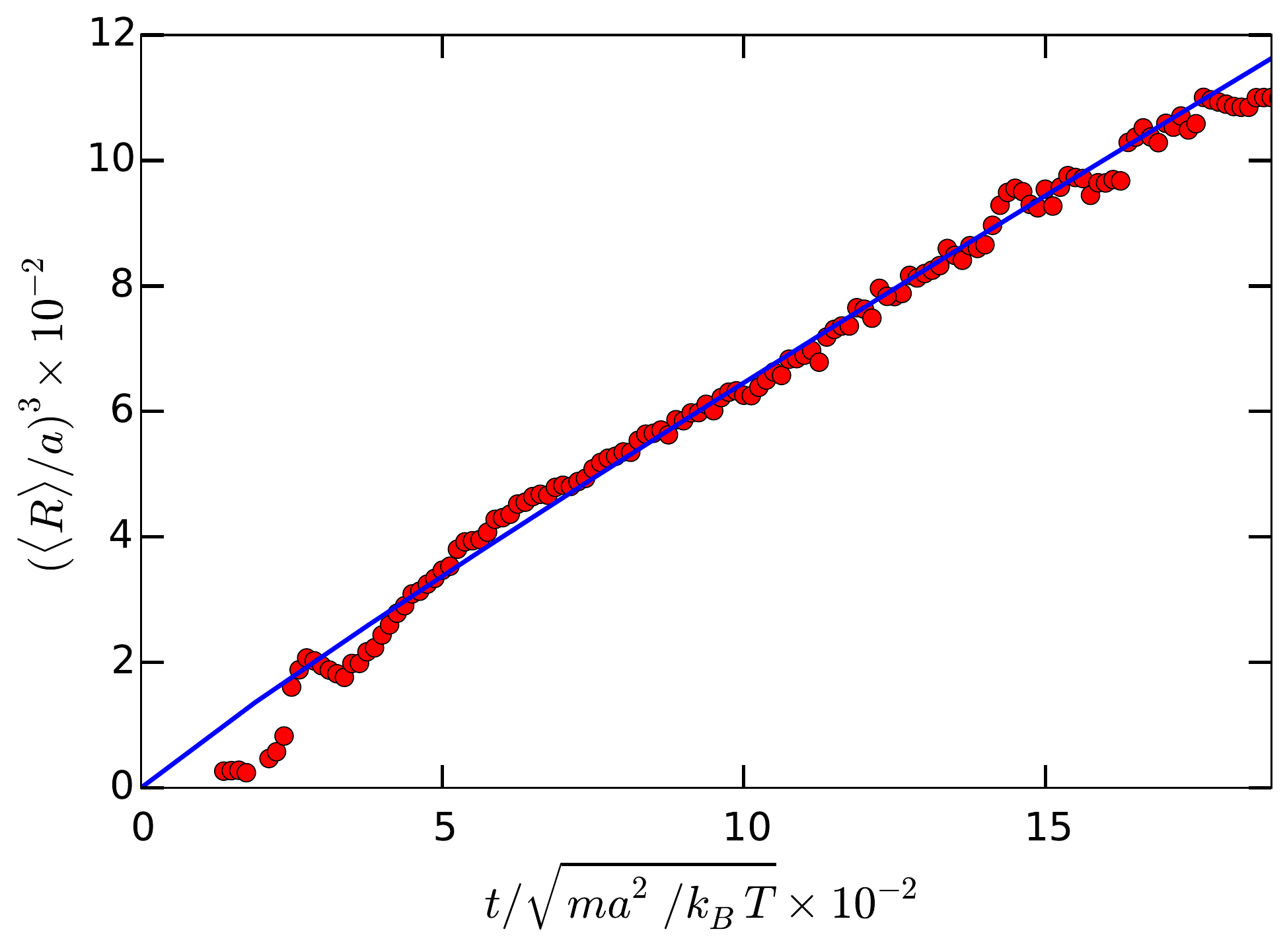}
  \caption{(Color online) Average radius (volume) of droplets (3D) as function of time. The simulation parameters are $\tau\,  m^{2}/(k_B T a^3)=-0.25$, $b  \,  m^{4} /(k_B T a^9) =10^{-3}$, $\bar{\chi}\, m/ k_B T =5$, $\kappa \, m^2/(k_B T a^5) = 0.3$, and the box size is $L_x=L_y=L_z=130a$. The solid line corresponds to the fit of Eq.~(\ref{ostwald}) in the interval $t/\sqrt{ma^2/k_B T} \in [300,\, 1500]$. }
  \label{ost_fig}
\end{center}
\end{figure}

{\em Droplet growth---Ostwald ripening.}
The generalization of the algorithm to three dimensions is straightforward. To illustrate the time
evolution of a phase separating system, we employ  the  Ginzburg-Landau parameters
$\tau  m^{2}/(k_B T a^3)=-0.25$, $b  m^{4} / (k_B T a^9) =10^{-3}$, $\kappa \,  m^{2}/(k_B T a^5)  = 0.3$,
$\bar{\chi}\,  m/k_B T =5$, and the particle fraction $N_A/N_B = 4$.
Similar to the 2D simulations, the actual measured temperature increases slightly by about $3\%$.
Figure~\ref{3D_1} presents three snapshots at different simulation times.
Initially, the system is totally mixed and evolves to the minimal free-energy conformation.
Over time, critical nuclei appear, grow, and coalesce to form larger clusters and droplets. The growth
of the clusters is expected to be described by the time evolution of the average radius (Ostwald ripening \cite{bald:02})
\begin{equation}
\langle R \rangle^3 \sim t^\epsilon \, ,
\label{ostwald}
\end{equation}
where the exponent $\epsilon=1$ in the diffusive regime. Figure~\ref{ost_fig} shows $\langle R \rangle ^3 $ as a function of time. After an initial time, $\langle R \rangle^3$ increases approximately in a linear manner. The fit of  Eq.~(\ref{ostwald}) yields the exponent $\epsilon\approx 0.94 \pm 0.01$, in close agreement with the theoretical value for diffusive growth.

\subsection{Interfacial width}

In equilibrium, the order-parameter profile of a flat interface  in the framework of the Landau theory is
given by
\begin{equation} \label{eq:tanh_MPC}
\phi= \phi_b \, \tanh\left( z/\xi_0 \right) \, ,
\end{equation}
where $\phi_b$ is the bulk order parameter, $z$ is the coordinate normal to the interface,
and the correlation length $\xi_0 = \left( -\kappa/2\tau  \right)^{1/2}$ characterizes the interfacial
width.

\begin{figure}[t]
\begin{center}
\includegraphics[width=0.9\columnwidth]{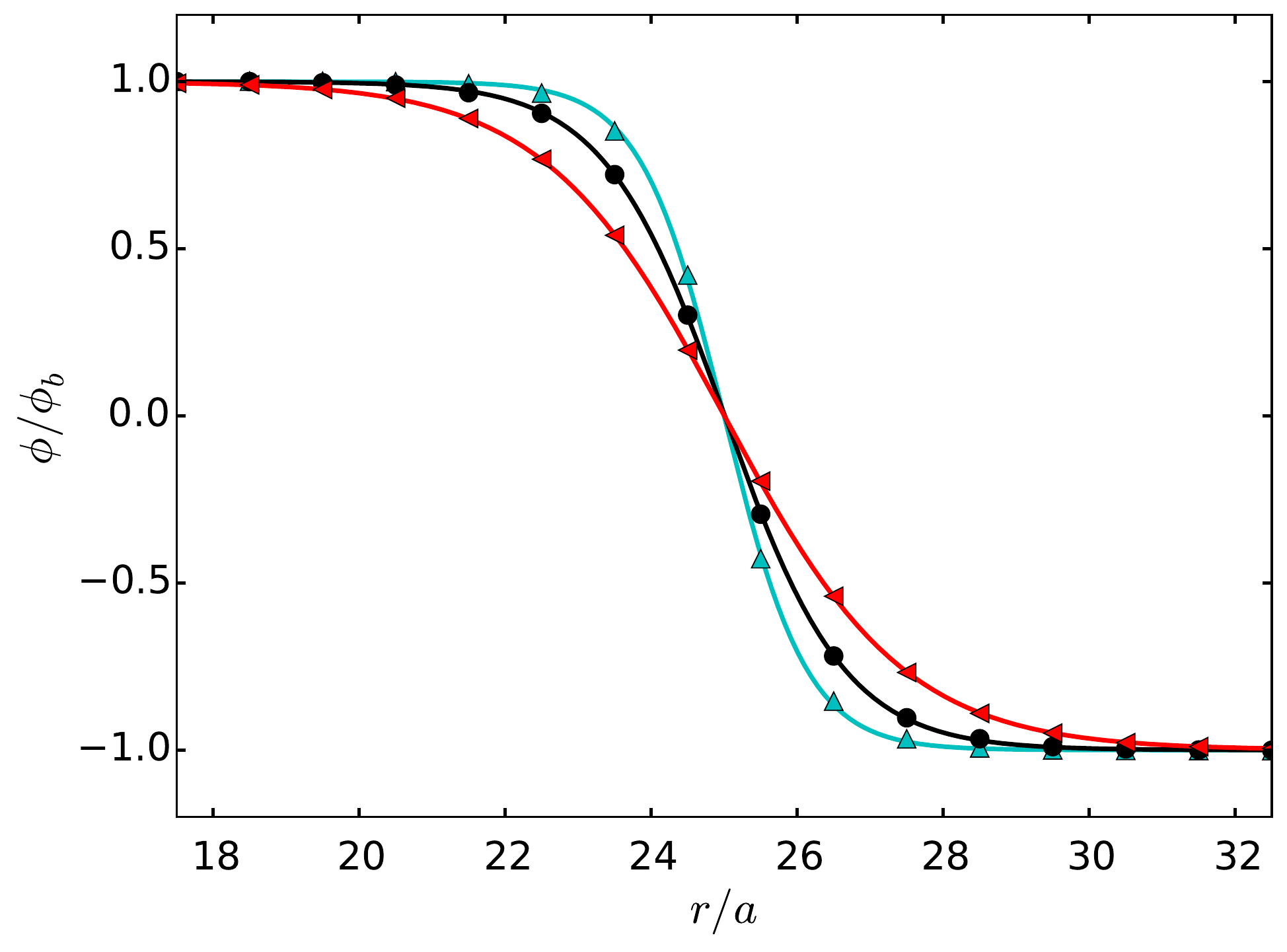}
  \caption{(Color online) Interfacial order-parameter profiles of a flat interface (3D) for  $\kappa = 0.3$ (cyan),
$0.6$ (black), and $1.2\,  k_B T a^5 m^{-2}$ (red). The other parameters are
$\tau\, m^2/(k_B T a^3)=-0.25 $, $b \, m^4/( k_B T a^9)=10^{-3}$, and $\bar{\chi}\, m/k_B T =5 $.
Fits (lines) of Eq.~(\ref{eq:tanh_MPC}) yield interfacial widths of
$\xi_0 / a= 1.14 $, $\xi_0/a = 1.65 $, and $\xi_0/a = 2.47 $.}
  \label{3D_2}
\end{center}
\end{figure}

To determine the interface profile, we fill the two halfs of the simulation box with $A$ and $B$
particles, respectively. The Ginzburg-Landau and MPC parameters are chosen to be the same as above
(3D system), except $\kappa$ is varied from $0.3$, $0.6$, to $1.2\,  k_B T a^5 m^{-2}$. The obtained
stationary state profiles are shown in Fig.~\ref{3D_2}. Fits to Eq.~(\ref{eq:tanh_MPC}) yield the interfacial
widths $\xi_0/a = 1.14 $, $1.65 $, and $2.47 $, whereas Ginzburg-Landau theory predicts
$\xi_0 /a= 0.77 $, $ 1.10 $, and $1.55 $. Hence, the simulation values are approximately $50\%$ larger
than those predicted by the Ginzburg-Landau theory. This quantitative deviation is not surprising, since
the MPC algorithm includes thermal fluctuations, which are not taken into account by the Ginzburg-Landau approach.
Indeed, the correlation length increases as $\xi_0\sim |\tau-\tau_c|^{-\nu}$ at the critical
temperature, where $\nu=0.5$ in Ginzburg-Landau theory, but $\nu=0.63$ for the three-dimensional
Ising model. Thus, fluctuations are expected to increase the interface width, in good qualitative
agreement with our simulation results.
Moreover, the interface widths are on the order of the size of a collision cell, where discretization
effects become visible; this makes a detailed quantitative comparison difficult.

\section{Conclusions}

We have proposed a thermodynamically consistent extension of the MPC algorithm for binary-fluid mixtures,
by coupling the Ginzburg-Landau free-energy functional of the density difference between the two fluids to
the particle motion in the MPC streaming step. The fluids phase-separate and form, depending on the
relative concentration, flat or curved (circle (2D), sphere (3D)) interfaces. The numerical results
reproduce the expected capillary-wave spectrum and fulfill the Young-Laplace pressure relation.
The analysis of the time dependence of phase separation in 3D,  i.e., of the Ostwald-ripening process,
proceeds with an exponent $0.94$, very close to the theoretically expected value (Lifshitz-Slyozov-Wagner
growth) in the diffusive regime.

A major advantage of the proposed augmented description of the multicomponent system with an auxiliary
order-parameter field is that a broad spectrum of nonlinear fluid equations of state can be implemented
easily and conveniently in an otherwise ideal-gas-type fluid.
This makes the MPC approach to hydrodynamics applicable to a wide range of phenomena in the
dynamics of complex fluids.  This approach is particularly useful for studies of synthetic phoretically
driven microswimmers, where various kinds of fluids are essential for propulsion \cite{elge:15,bech:16}.

There are various aspects of our novel technique, which deserve further and more detailed study. On the one hand, additional properties of the binary MPC fluid should be addressed, such as transport coefficients or correlation functions. On the other hand, dynamical quantities and critical behavior can be studied. Moreover, extensions are possible to explore the dynamical properties of binary fluids adjacent to walls. Since the free-energy functional of confined systems is available, studies of  wetting behavior and contact angles are within reach.

\section{Appendix A: Discretization scheme}
The fluid velocity obeys the Navier-Stokes equation
\begin{equation}
\rho \left( \partial_t v_\alpha + v_\beta \partial_\beta v_\alpha \right) = \eta \partial_\beta \partial_\beta v_\alpha - \partial_\beta \mathcal{P_{\alpha \beta}}. \label{forceT1}
\end{equation}
Here, the Einstein summation convention over the spatial components $\alpha$, $\beta$ is used. The pressure tensor comprises contributions from the standard MPC algorithm as well as those of the Ginsburg-Landau free-energy functional. The latter contribution can, on the collision cell-level, be equally well written as body-force density acting on the fluid
\begin{align}
\partial_{\beta} \mathcal{P}_{\alpha \beta}^{GL}  = \rho \partial_{\alpha} \frac{\delta \mathcal{F}}{ \delta \rho} + \phi \partial_{\alpha} \frac{\delta \mathcal{F}}{ \delta \phi}  = -f_{\alpha} \label{discrete2}.
\end{align}
The change of the velocity due to transport of the immiscible fluids is \cite{thei:99} $\rho \dot{v}_\alpha = -\phi \partial_\alpha \left( \delta \mathcal{F} / \delta\phi \right)= -\partial_\beta  \left( \mathcal{P}_{\alpha \beta}^{b} + \mathcal{P}_{\alpha \beta}^{int} \right)$.
Considering particles of type $A$ in the discretization volume, i.e., the collision cell, this expression can be written as
\begin{equation}
\dot{v}_\alpha^A = - \partial_\alpha \frac{\delta \mathcal{F}}{\delta\phi} = -\frac{\partial_\beta \mathcal{P}_{\alpha \beta}^{b}}{\phi} + \kappa \partial_\alpha \left( \nabla^2 \phi \right).
\label{discrete}
\end{equation}
We discretize the first term on the RHS of (\ref{discrete}) by averaging over two neighbouring cells
\begin{align}
& \frac{\partial_x \mathcal{P}_{x x}^{b}}{\phi}   \approx \frac{ \tau \left( \phi^2_{x+1,y} -\phi^2_{x,y} \right) + \frac{3}{2} b  \left( \phi^4_{x+1,y} - \phi^4_{x,y} \right)}{\phi_{x+1,y} + \phi_{x,y}  } \\ \nonumber
& =  \tau \left( \phi_{x+1, y} - \phi_{x, y} \right)
 + \frac{3b}{2} \left( \phi_{x+1, y}^2 + \phi_{x, y}^2 \right) \left(  \phi_{x+1, y} - \phi_{x, y} \right),
\end{align}
which prevents a particle drift.
Here, the force acting on the edges of a collision cell. Hence, when distributing this force onto the particles, the cell $(x,y)$ as well as neighbouring cells (i.e., $(x+1,y)$ or $(x,y+1)$, depending on the Cartesian component) have to be considered.
For the second term in (\ref{discrete}), we simply choose a central difference scheme. The discretization scheme for the particles of type $B$ differs just by a minus sign in the ``bulk'' and ``interfacal'' contributions. \\
The non-linear pressure contribution is discretized in the same way as the ``bulk'' term above, with the only difference that no distinction between type $A$ or $B$ particles has to be made.


\acknowledgments

We thank A. Wysocki for helpful discussions and suggestions. Support of this work by the DFG priority program SPP 1726
on ``Microswimmers -- from Single Particle Motion to Collective Behaviour'' is gratefully acknowledged.

\bibliographystyle{eplbib}

\end{document}